\documentclass[12pt]{iopart}
\usepackage{graphicx}

\begin{document}

\title[Y.G. Ma et al. ]{$\Delta$-scaling
 and heat capacity in relativistic ion collisions}

\author{Y. G. Ma\dag\footnote[1]{To
whom correspondence should be addressed: ygma@sinap.ac.cn}, G. L.
Ma\dag\ddag, X. Z. Cai\dag, J. G. Chen\dag, J. H. Chen\dag, D. Q.
Fang\dag, W. Guo\dag, Z. J. He\dag, H. Z. Huang\S, J. L. Long\dag,
C. W. Ma\dag, B. H. Sa$\|$, W. Q. Shen\dag, Q. M. Su\dag, K.
Wang\dag,\\ Y. B. Wei\dag, T. Z. Yan\dag, C. Zhong\dag, J. X.
Zuo\dag }

\address{\dag\ Shanghai Institute of Applied Physics, Chinese Academy of Sciences,  China}
\address{\ddag\ Graduate School of the Chinese Academy of Sciences, China}
\address{\S Dept of Physics and Astronomy, University of California at Los
Angeles,  USA}
\address{$\|$ China Institute of Atomic Energy, P.O.Box 918, Beijing 102413, China}

\begin{abstract}
The  $\Delta$-scaling method has been applied to the total
multiplicity distribution of the relativistic ion collisions of
p+p, C+C and Pb+Pb which were simulated by a Monte Carlo package,
LUCIAE 3.0. It is found that the $\Delta$-scaling parameter
decreases with the increasing of the system size. Moreover, the
heat capacities of different mesons and baryons have been
extracted from the event-by-event temperature fluctuation in the
region of low transverse mass and they show the dropping trend
with the increasing of impact parameter.
\end{abstract}



%

Studying fluctuation and correlation is an important issue to
investigate matter properties formed in relativistic
nucleus-nucleus collisions. Some observables have been suggested
to quantify the fluctuation, such as the balance function, net
charge fluctuation, multiplicity fluctuation, transverse momentum
fluctuation, particle ratio fluctuation and $\Phi_{PT}$ etc
\cite{Fluct}. In this article we will discuss some novel methods.
We will apply a universal fluctuation method ($\Delta$-scaling)
which was recently proposed to explore the order-disorder phase
transition in the low-intermediate energy heavy ion collisions
\cite{scaling,Ma} to discuss multiplicity fluctuation in
relativistic ion collisions. In addition, the heat capacity of
hadrons are extracted from the event-by-event temperature
fluctuation.

$\Delta$-scaling is observed when two or more probability
distributions $P[m]$ of the stochastic observable $m$ collapse
onto a single scaling curve $\Phi$(z) if a new scaling observable
is defined by
 $   z  = \frac{(m-m^*)}{\langle m\rangle^\Delta}$.
This curve is \cite{scaling}:
\begin{equation}
\langle m\rangle^\Delta P[m] =  \Phi (z) \equiv \Phi
[\frac{m-m^*}{\langle m\rangle^\Delta}]
\end{equation}
where  $\Delta$ is a scaling parameter, $m^*$ is the most probable
value of $m$, and $\langle m\rangle$ is the mean of $m$. If we
assume that $P[m]$ is a Gaussian distribution, we have $P [m] =
\frac{1}{\sigma \sqrt{2\pi}} exp[-\frac{1}{2}
(\frac{m-\mu}{\sigma})^2  ]$,
 where $\mu = \langle m\rangle = m^*$, $\sigma$ is the width of
the Gaussian distribution, both depending on incident energy. If
this Gaussian distribution $P[m]$ obeys $\Delta$-scaling law, we
have
\begin{equation}
   \sigma  \propto \mu^\Delta  .
\end{equation}

In this work, p+p, C+C and Pb+Pb in SPS energies (20-200 AGeV)
have been investigated with the help of LUCIAE3.0 model
\cite{Luciae}. The head-on (b=0fm) collisions are simulated in
this work. The LUCIAE is an extension Monte Carlo model of the
FRITIOF \cite{Fritiof}, in which a nucleus - nucleus collision is
described as the independent  sum of nucleon-nucleon collisions.
For more details of LUCIAE, please see Ref. \cite{Luciae}.

\begin{figure}
\vspace{0.2truein}
\includegraphics[scale=.5]{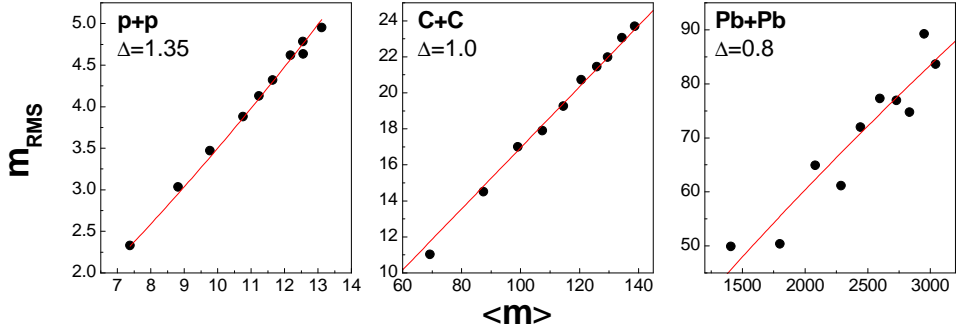}
\vspace{-2.6truein} \caption{The determination of the
$\Delta$-scaling parameter using the power-law fits to the
relationship $m_{RMS}$ vs $\langle m\rangle$.  From the left to
right, it corresponds to p + p with $\Delta$ = 1.35, C + C with
$\Delta$ = 1.00 and Pb+Pb with $\Delta$ = 0.80. The circles are
the calculated values and the lines represent the power-law fits.}
\label{Fig_fix}
\end{figure}

In order to better investigate the $\Delta$-scaling, we can use
the power-law fit to the relationship between the RMS width of the
multiplicity distribution ($m_{RMS}$) and the mean value of the
multiplicity distribution ($\langle m\rangle$) according to
Eq.(1). If there exists a good fit: $m_{RMS} = c_0 \langle
m\rangle ^\Delta$ where $c_0$ is a fit parameter, indicating  that
a good $\Delta$-scaling law is satisfied with a  scaling parameter
 $\Delta$.
In this way, we found that the best fit value of $\Delta$ is 1.35
for p+p, 1.00 for C+C and 0.80 for Pb + Pb, respectively, which is
shown in Figure 1. In the figure, each point represents an energy
point from 20 AGeV to 200 AGeV with the interval of 20 AGeV. When
$\Delta$ = 1, the RMS width (fluctuation) is proportional to its
mean value. This is the case of C+C system. For light system p+p,
the rising of the RMS width is  faster than its mean value with
the increasing of beam energy. In contrary, for heavier system,
the growing of the RMS width is slower than the mean value. In all
three cases, $\Delta$-value is larger than 1/2, which is the case
of the independent particle emission (Poisson distribution). The
decreasing behaviour of $\Delta$-value with the system size
reflects how the multiplicity fluctuation grows with its mean
value. Its physics origin may stem from the stronger rescattering
effect in larger system which induce some correlations in
multiparticle production process and in consequence the
fluctuation becomes weaken.

 Using these values of $\Delta$, the multiplicity distributions
can be compressed onto a unique curve for a given collision system
from 20 AGeV to 200 AGeV. Figure 2 shows the $\Delta$-scaling
curves for p+p, C+C and Pb+Pb systems with $\Delta$ value of 1.35,
1.00 and 0.80, respectively.

\begin{figure}
\begin{center}
\vspace{0.1truein}
\includegraphics[scale=.5]{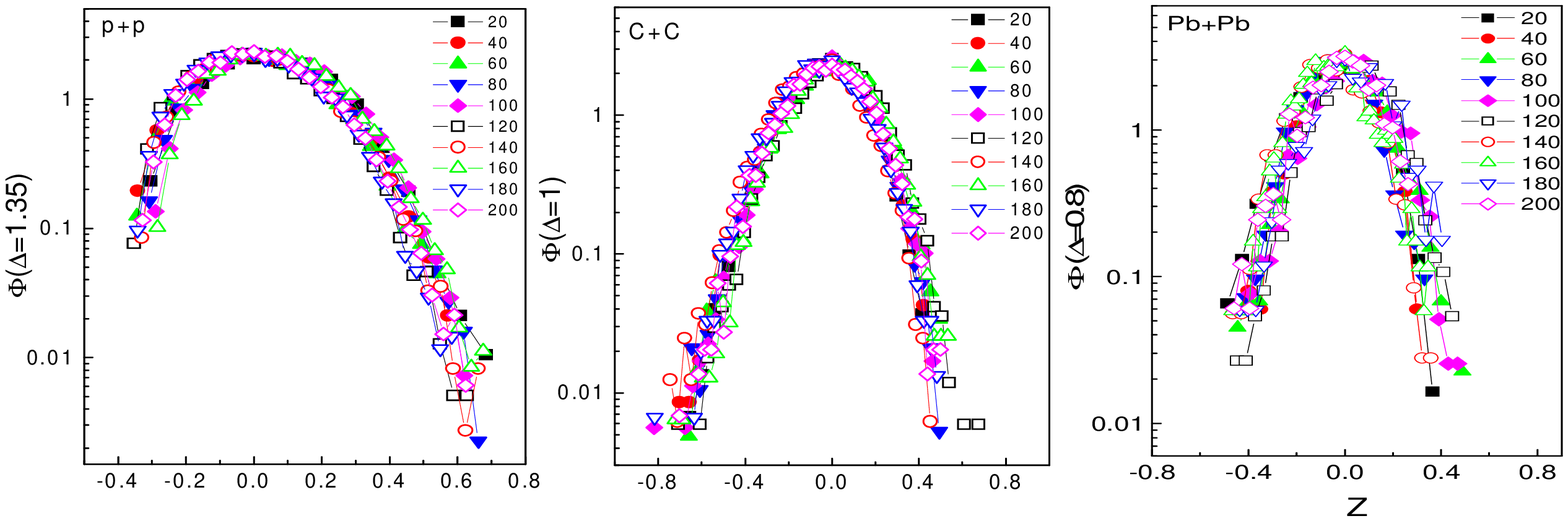}
\end{center}
\vspace{-2.7truein} \caption{$\Delta$-scaling for p + p with
$\Delta$ = 1.35, for  C + C with $\Delta$ = 1.00 and for Pb+Pb
with $\Delta$ = 0.80 for the different beam energies.}
\label{Fig_dens}
\end{figure}

In relativistic collisions, the transverse mass  ($m_T$)
distribution of the particles in lower $m_T$ region can be
approximately  described as $ \frac{1}{m_T} \frac{dN}{d m_T}  = A
e^ {-\frac{m_T}{T}}$ where $A$ is a normalized coefficient which
is related to the volume/multiplicity term and $T$ is an apparent
temperature. Thanks to the large enough particle multiplicity, it
is possible to extract $A$ and $T$ for particles on event-by-event
basis. In this way, the event-by-event temperature distribution
can be constructed. Usually, such kind of temperature distribution
($P(T)$) can be described by \cite{Landau,Shur}
\begin{equation}
P(T) \sim exp[-C_v(\frac{\Delta T}{T})^2 ],
 \end{equation}
where $\Delta T$ is the deviation ($T-\langle T\rangle$) of
temperature from the mean value ($\langle T\rangle$) and $C_v$ can
be explained as the heat capacity of a certain hadron. Since the
heat capacity is an extensive observable, we define a normalized
heat capacity $C_v/N$, i.e. the heat capacity per hadron
multiplicity ($N$) which corresponds to the energy that is
required to go up one unit temperature for producing one particle.

The heat capacity of different particles (in the range of $m_T-m_0
< 0.5$ GeV) has been extracted by Eq.(3). The left panel of Figure
3 shows an example of event-by-event temperature distribution of
$\pi^+$ for the head-on Pb+Pb collisions at 160 AGeV. From this
kind of distribution, we apply Eq.(3) to extract the heat capacity
$C_v$.  The curve of the figure depicts this fit. In this way, we
get the heat capacity for various hadrons in different energies.
The middle panel of Figure 3 gives the dependence of heat capacity
of various particles ($\pi^+$, $k^+$, p and $\Lambda$) on incident
energy in head-on Pb+Pb collision. From this figure, the
normalized heat capacity of various particles decreases with the
increasing of incident energy and tends to a saturation,
indicating that less energy is needed to rise the same temperature
in higher energies. In the viewpoint of fluctuation, this means
that the increasing $m_T$ fluctuation at higher energies. The
normalized heat capacity of the particles increases with mass of
particle, which reflects that more energy is needed to rise the
same temperature for heavier particles. In addition, the impact
parameter dependence of the heat capacity is also investigated for
$\pi^+$. The right panel of Figure 3 depicts the dependence of
$C_v/N$ for $\pi^+$ on the impact parameter for Pb + Pb. The
decreasing trend of $C_v/N$ indicates that the temperature
fluctuation becomes larger in peripheral collisions, which is
similar to the $P_T$ fluctuation or balance function as a function
of centrality in STAR data \cite{westfall}.

\begin{figure}
\begin{center}
\vspace{.1truein}
\includegraphics[scale=.45]{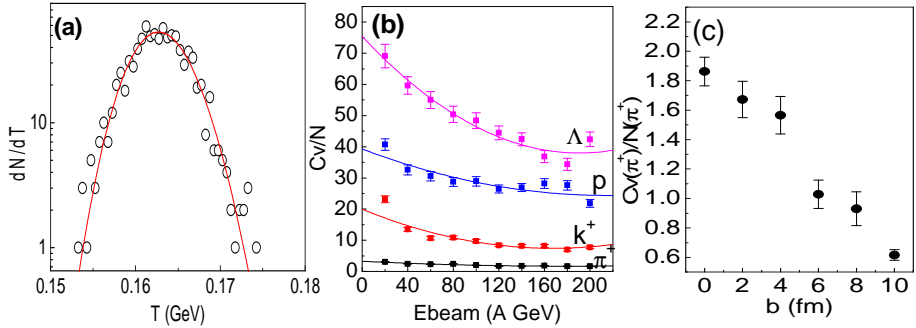}
\vspace{-2.2truein}
\end{center} \caption{(a) The event-by-event temperature distribution of $\pi^+$
in the head-on Pb+Pb collision at 160AGeV. The curve is a fit with
Eq.(3); (b) The normalized heat capacity of different mesons and
baryons as a function of incident energy for the head-on Pb + Pb
collisions. The lines represent the second polynomial fits to
guide the eyes; (c) The dependence of the normalized heat capacity
of $\pi^+$ on impact parameter for 160 AGeV Pb + Pb collisions.
The error bars are statistical.} \label{fig_Cv-PID}
\end{figure}

In summary, we have demonstrated that there exists the
$\Delta$-scaling for the total multiplicity of charged particles
for the simulated head-on collisions of p + p, C + C and Pb + Pb
from $E_{lab}$ = 20 to 200 AGeV in LUCIAE model. That means that
the multiplicity distributions obey a certain kind of universal
laws, regardless of beam energies and collision systems. It is
found that the scaling value of the $\Delta$ decreases with the
system size, which reflects that the growth of the fluctuation
with the multiplicity is faster in light system. The
event-by-event thermal fluctuation in lower $m_T$ region is
constructed and the normalized heat capacities of the different
mesons and baryons are extracted. It is found that the heat
capacity per hadron decreases with the increasing incident energy
and impact parameter, while it increases with mass of particle.
Considering that the LUCIAE model is a string-hadronic model, the
partonic effect which becomes more important in RHIC energies is
absent in the model. This effect should be investigated in near
future. The work along this line is in progress.

This work was partially supported by the Major State Basic
Research Development Program under Contract No G2000774004, the
National Natural Science Foundation of China (NNSFC) under Grant
No 10328259 and 10135030.

{}


\section*{References}

\begin{thebibliography}{}
\bibitem{Fluct}Proceeding of Quark Matter 2004, {\it J. Phys.} G {\bf 30}, edited by H. G.
Ritter and X. N. Wang.
\bibitem{scaling}  Botet R {\it et al.} 2001 {\it Phys. Rev. Lett.} ${\bf 86}$  3514
\bibitem{Ma} Ma Y G {\it et al.} 2004 {\it ArXiv:Nucl-ex/0410018},
submitted to {\it Phys. Rev.} C.
\bibitem{Luciae} Tai A and Sa B H 1999 {\it Comp. Phys. Commu.} ${\bf 116}$ 355
\bibitem{Fritiof}  Pi H  1992 {\it Comp. Phys. Comm.} {\bf 71} 173
\bibitem{Landau} Laudau L D and Lifschitz I M {\it Course of Theoretical Physics: Statistical Physics}
Vol.5.
\bibitem{Shur} Shuryak E V 1998 {\it Phys. Lett.} B {\bf 423} 9.
\bibitem{westfall} Westfall G D 2004 {\it J. Phys.} G {\bf 30}
S1389.
\end{thebibliography}
\end{document}